\documentclass[12pt]{iopart}

\usepackage{iopams}
\usepackage{hyperref}
\usepackage{cite}
\hypersetup{hypertex=true,colorlinks,linkcolor=cyan,anchorcolor=cyan,citecolor=cyan,filecolor=cyan,urlcolor=cyan}
\usepackage{graphicx}
\usepackage{bm}
\usepackage[utf8]{inputenc}
\usepackage[T1]{fontenc}
\usepackage[left]{lineno}
\usepackage{xcolor}
\begin{document}
\title{Diagnosis of Fast Electron Transport by Coherent Transition Radiation}

\author{Yangchun Liu$^1$, Xiaochuan Ning$^1$, Dong Wu$^{2,*}$, Tianyi Liang$^1$, Peng Liu$^1$, Shujun Liu$^1$, Xu Liu$^2$, Zhengmao Sheng$^{1,*}$, Wei Hong$^3$, Yuqiu Gu$^3$, Xiantu He$^1$}

\address{$^1$Institute for Fusion Theory and Simulation, School of Physics, Zhejiang University, Hangzhou, 310058, China}
\address{$^2$Key Laboratory for Laser Plasmas and School of Physics and Astronomy, and Collaborative Innovation Center of IFSA (CICIFSA), Shanghai Jiao Tong University, Shanghai, 200240, China}
\address{$^3$Science and Technology on Plasma Physics Laboratory, Research Center of Laser Fusion, China Academy of Engineering Physics, Mianyang, 621900, China}

\ead{\mailto{dwu.phys@sjtu.edu.cn},\mailto{zmsheng@zju.edu.cn}}

\vspace{10pt}
\begin{indented}
\item[]\today
\end{indented}

\begin{abstract}
Transport of fast electron in overdense plasmas is of key importance in high energy density physics. However, it is challenging to diagnose the fast electron transport in experiments. In this article, we study coherent transition radiation (CTR) generated by fast electrons on the back surface of the target by using 2D and 3D first-principle particle-in-cell (PIC) simulations. In our simulations, {aluminium target of $2.7$ g/cc is simulated in two different situations by using a newly developed high order implicit PIC code. Comparing realistic simulations containing collision and ionization effects, artificial simulations without taking collision and ionization effects into account significantly underestimate the energy loss of electron beam when transporting in the target, which fail to describe the complete characteristics of CTR produced by electron beam on the back surface of the target. Realistic simulations indicate the diameter of CTR increases when the thickness of the target is increased. }
This is attributed to synergetic energy losses of high flux fast electrons due to Ohm heatings and colliding drags, which appear quite significant even when the thickness of the solid target only differs by micrometers. Especially, when the diagnosing position is fixed, we find that the intensity distribution of the CTR is also a function of time, with the diameter increased with time. As the diameter of CTR is related to the speed of electrons passing through the back surface of the target, our finding may be used as a new tool to diagnose the electron energy spectra near the surface of solid density plasmas.
\end{abstract}

%
%
\submitto{\NJP}
%
%
%

\section{Introduction}
With the invention of chirped pulse amplification technology\cite{Mourou2006RMP}, the laser intensity increased by six orders of magnitude, and the research work in the field of high-energy density physics has been effectively promoted. When the laser intensity is higher than $10^{13} \ \mathrm{W/cm^2}$, the laser will ionize the target material rapidly by multiphoton ionization, tunneling ionization and so on, forming plasma composed of electrons and ions. When the laser intensity reaches $10^{18} \ \mathrm{W/cm^2}$, it will cause relativistic oscillation of electrons in the laser field and heat the temperature of electrons to more than $1 \ \mathrm{MeV}$. Such laser can also be used to accelerate high-energy charged particles, produce ultra-short and ultra-strong gamma rays and even positive and negative electron pairs. One of the key subjects in high energy density physics is the transport of fast electrons. It has many potential applications in various fields, such as laser accelerators\cite{Tajima1979PRL,Modena1995Nature}, fast ignition\cite{Tabak1994POP,Kodama2001Nature}, positron–electron plasmas \cite{Edison1998PRL,Cowan1999LPB} and terahertz source\cite{Liao2016PRL,Liao2017PPCF}. In experiments, $K\alpha$, extreme ultraviolet (XUV) emission\cite{Stephens2004PRE,Lancaster2007PRL}, shadowgraphy \cite{Tatarakis1998PRL,Gremillet1999PRL} and optical emission \cite{Santos2002PRL,Baton2003PRL} are usually used to study fast electron transport.
In particular, coherent transition radiation (CTR) is becoming an effective tool to diagnose the fast electron population,
especially for those electrons micro-bunched in time\cite{Baton2003PRL,Zhengjian2004PRL,Huang2016POP,Popescu2005POP,Blakeney2020POP,Santos2002PRL,Dechard2020PoP,Ding2016PRE,Hu2020PRA,Mondal2020PRApplied}.
\par

Transition radiation is the radiation emitted by charged particles passing through the interface between two media with different
dielectric constants. The electron beam is periodically produced by different laser heating mechanisms such as vacuum heating, resonance absorption and $\textbf{J}\times \textbf{B}$ heating.
Therefore, the transition radiation generated by such electron beam through the interface is easily coherent. It makes the CTR spectra peak at $n\omega_0$\cite{Zhengjian2003POP}, where $n$ is a positive integer and $\omega_0$ is laser frequency.
We can infer the main laser heating mechanism of electrons by diagnosing CTR. In addition, CTR is also a potential scheme for intense terahertz radiation sources\cite{Liao2016PRL}.
\par
In this article, we study the CTR generated by fast electrons on the back surface of the target with 2D and 3D particle-in-cell (PIC) simulations.
{Realistic simulations containing collision and ionization effects indicate the diameter of CTR increases when the thickness of the target is increased.} This is attributed to synergetic energy losses of high flux fast electrons due to Ohm heatings and colliding drags \cite{Kemp2006PRL,Sherlock2014PRL} which appear quite significant even when the thickness of the solid target only differs by micrometers.
Especially, when the diagnosing position is fixed, we find that the intensity distribution of the CTR is also a function of time, with the diameter increased with time. As the diameter of CTR is related to the speed of electrons passing through the back surface of the target, our finding may be used as a new tool to diagnose the electron energy spectra near the surface of solid density plasmas. {Meanwhile, in order to understand the role played by collision and ionization physics, artificial simulations without taking into collision and ionization effects are also performed. When comparing with realistic simulations, artificial simulations significantly underestimate the energy loss of electron beam when transporting in the target, which unable to describe the complete characteristics of CTR produced by electron beam on the back surface of the target.}

\section{Theoretical Model}

\begin{figure}[!t]
	\centering	
	\includegraphics[scale=0.65]{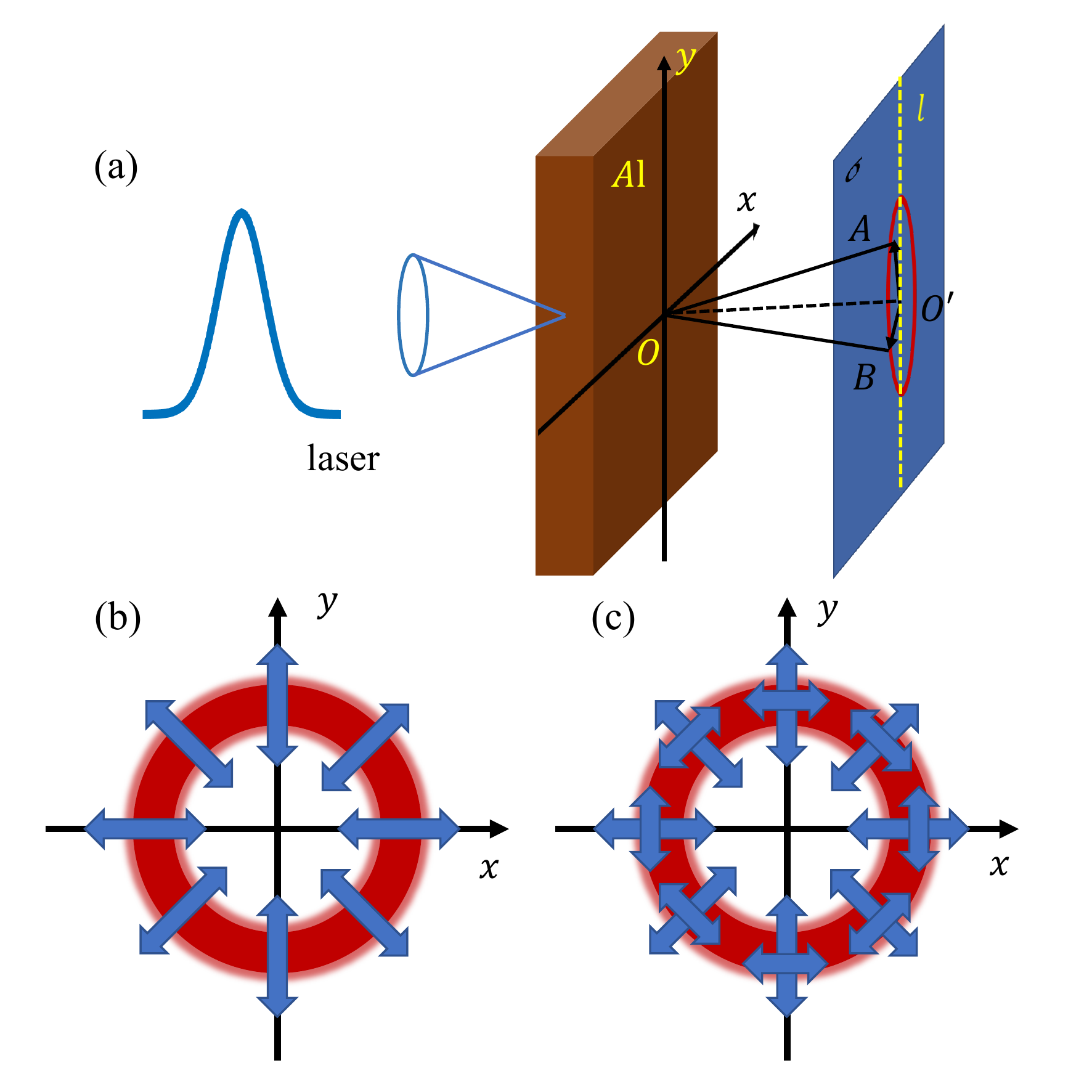}
	\caption{(a) Schematic diagram of PIC simulation. (b), (c) is the polarization characteristics of CTR for different electron beams vertically and obliquely passing through the interfaces, respectively.}
	\label{fig1}
\end{figure}

\begin{figure}[!t]
	\centering
	\includegraphics[scale=0.8]{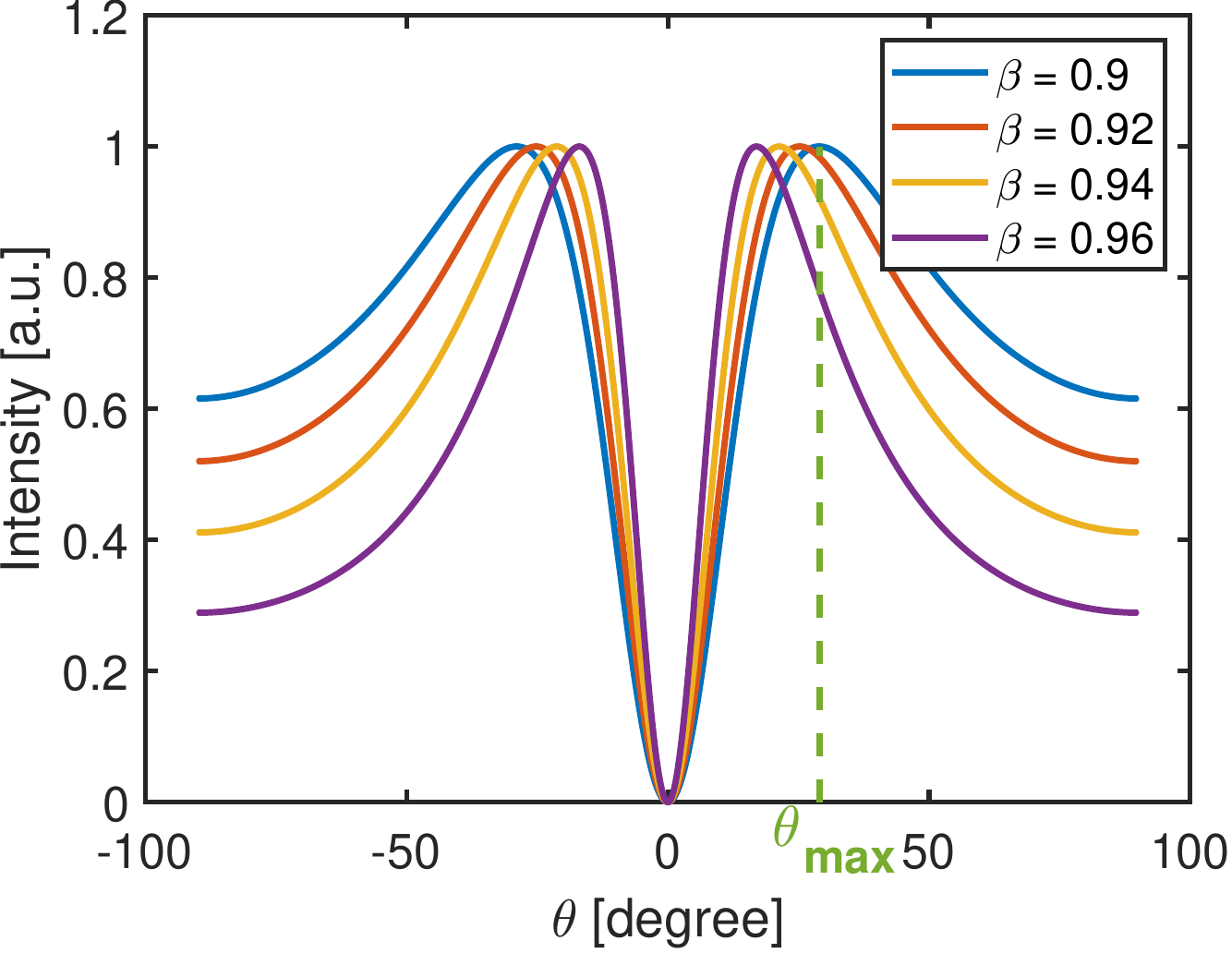}	
	\caption{Normalized intensity distribution of CTR when an electron beam vertically passes through the interface. Different colors correspond to different electron energies, $\beta=v/c$.}
	\label{fig2}
\end{figure}
Here, we need to emphasize several properties of CTR. The first characteristic of CTR is polarization. The polarization characteristic of CTR is called radial polarization in optics\cite{Deng2007OL}. The polarization component of a certain CTR ray lies in the plane formed by the ray and the main direction. As shown in figure\ \ref{fig1}(a), the polarization direction of ray $OA$ is in plane $OO'A$, where $OO'$ is the main direction of radiation. That is to say, for the electron beam vertically passing through the interface, the projection of radiation electric field on $XOY$ plane is along the radial direction, as shown in figure\ \ref{fig1}(b). For the electron beam obliquely passing through the interface, CTR is not completely radially polarized, as shown in figure \ref{fig1}(c). This electric field not only has components parallel to the radial direction, but also has components perpendicular to the radial direction, which was observed by Happek et al.\cite{Happek1991PRL}, the radiation angle distribution is similar, but the intensity is not completely symmetrical\cite{Pakluea2020}.

The second is characteristic of the angular distribution. Both CTR and incoherent transition radiation (ITR) follow the same characteristic of angular distributions, which satisfy the formula described in \cite{Zhengjian2003POP},
	\begin{eqnarray}
		\frac{\mathrm{d}\varepsilon _{\mathrm{single}}}{\mathrm{d}\omega \mathrm{d}\Omega}=\frac{e^2}{\pi ^2c}\frac{\beta ^2\cos \Theta \left[ \sin \theta -\beta \sin \Theta \cos \left( \phi -\Phi \right) \right] ^2}{\left[ \left( 1-\beta \sin \theta \sin \Theta \cos \left( \phi -\Phi \right) \right) ^2-\beta ^2\cos ^2\theta \cos ^2\Theta \right] ^2},
	\end{eqnarray}
where $(\theta,\phi)$ and $(\Theta,\Phi)$ represent the angle between Z and X axes and the direction for the radiation and the electron beam, respectively.
As shown in figure\ \ref{fig2}, when an electron beam vertically passes through the interface, the angular distribution of the transition radiation presents a conical distribution. The intensity of transition radiation in the direction of electron beam exit is zero, and there is a peak at a certain angle deviating from the direction of electron beam exit. This angle satisfies $\theta _{\mathrm{max}}=\mathrm{arcsin} ( \pm {(\left( \beta -1/\beta \right) ^2+\left( 1-\beta ^2 \right))^{1/2}} )$ and decreases with the increase of the electron beam energy\cite{Zhengjian2003POP}. That is to say, the faster the electron velocity passing through the rear surface of the target, the smaller the cone angle. In the relativistic limit, this angle is approximately equal to $1/\gamma$.

The third characteristic of CTR is reflected on the spectrum. Because of periodic laser heating mechanisms, transition radiation is always coherent. Resonance absorption and vacuum heating can heat and produce a electron bunch once in a cycle. This makes the transition radiation coherently superimposed at the integer multiple of the laser frequency. $\textbf{J}\times \textbf{B}$ heating can heat electrons twice in one laser cycle. This makes the transition radiation coherently superimposed at even times of laser frequency. Unde the combined action of various heat mechanisms, the spectrum of transition radiation will peak at $n\omega_0$, and the intensity will decrease exponentially with the increase of $n$\cite{Zhengjian2003POP}. Where $n$ is a positive integer and $\omega_0$ is laser frequency. Besides, the transition radiation in the low frequency range is always coherent. This can be seen from a simple model in \cite{Zhengjian2003POP},
\begin{equation}
	\frac{\mathrm{d}^2\varepsilon _{\mathrm{CTR}}}{\mathrm{d}\omega \mathrm{d}\Omega}=\left( N-1 \right) \left| \tilde{n}\left( \omega ,\boldsymbol{q} \right) \right|^2\frac{\mathrm{d}^2\varepsilon _{\mathrm{ITR}}}{\mathrm{d}\omega \mathrm{d}\Omega},
\end{equation}
with $\tilde{n}\left( \omega ,\boldsymbol{q} \right) =\tilde{n}_{\ell}\left( \omega \right) \tilde{n}_{\bot}\left( q \right)$,
\begin{equation}
	\tilde{n}_{\bot}\left( q \right) =\exp \left( -q^2a^2/2 \right) =\exp \left( -\frac{\omega ^2a^2}{2c^2}\sin ^2\theta \right) \nonumber
\end{equation}
and
\begin{eqnarray}
		\tilde{n}_{\ell}\left( \omega \right) =&&\frac{1}{1+\Delta \exp \left( -\omega _{0}^{2}\tau _{0}^{2}/2 \right)} \times \nonumber\\
		&&\left\{
		\exp \left( -\frac{\omega ^2\tau _{0}^{2}}{2} \right)
		\right.
		\left.+\frac{\Delta}{2}\exp \left( -\frac{\left( \omega -\omega _0 \right) ^2\tau _{0}^{2}}{2} \right) \right\},\nonumber	
\end{eqnarray}
where $\mathrm{d}^2\varepsilon _{\mathrm{ITR}}/\mathrm{d}\omega \mathrm{d}\Omega$ is the ITR unrelated to $\omega$, $N$ represents the number of electrons passing through the interface, $\tilde{n}\left( \omega ,\boldsymbol{q} \right)$ is Fourier transform of electron distribution in which $\boldsymbol{q}$ is the tangential radiation wave vector, $a$ is the beam radius, $\Delta$ is the microbunching amplitude, $\tau_{0}$ is the duration of electron pulse and $\omega_{0}$ is the microbunching frequency in the beam. According to the above formula, transition radiation is always coherent for $\omega \leqslant \tau _{0}^{-1}$. This makes CTR have high intensity in terahertz range\cite{Liao2016PRL}. Besides, in the study of Liu et al. and Batani et al., it is found that the intensity relationship between CTR and blackbody radiation is related to the thickness of the target\cite{Liu2018POP,Batani2006PPCF}. In a thin target, the optical coherent transition radiation has the same order of magnitude as the blackbody radiation, and may even be smaller than the blackbody radiation, but the CTR in terahertz range is stronger than that in blackbody. However, as the thickness of the target increases, the intensity of blackbody radiation decreases faster, so the optical coherent transition radiation dominates in thick targets. Because of the special polarization characteristics and angular distribution characteristics of CTR, the CTR for complete diagnosis should be in a ring shape. As shown by Pakluea' s plot of CTR intensity distribution according to theoretical formula\cite{Pakluea2020}.
\par

\section{Simulation Result}
\subsection{{Realistic simulations containing collision and ionization effects}}
\begin{figure}[!t]
	\centering
	\includegraphics[scale=0.6]{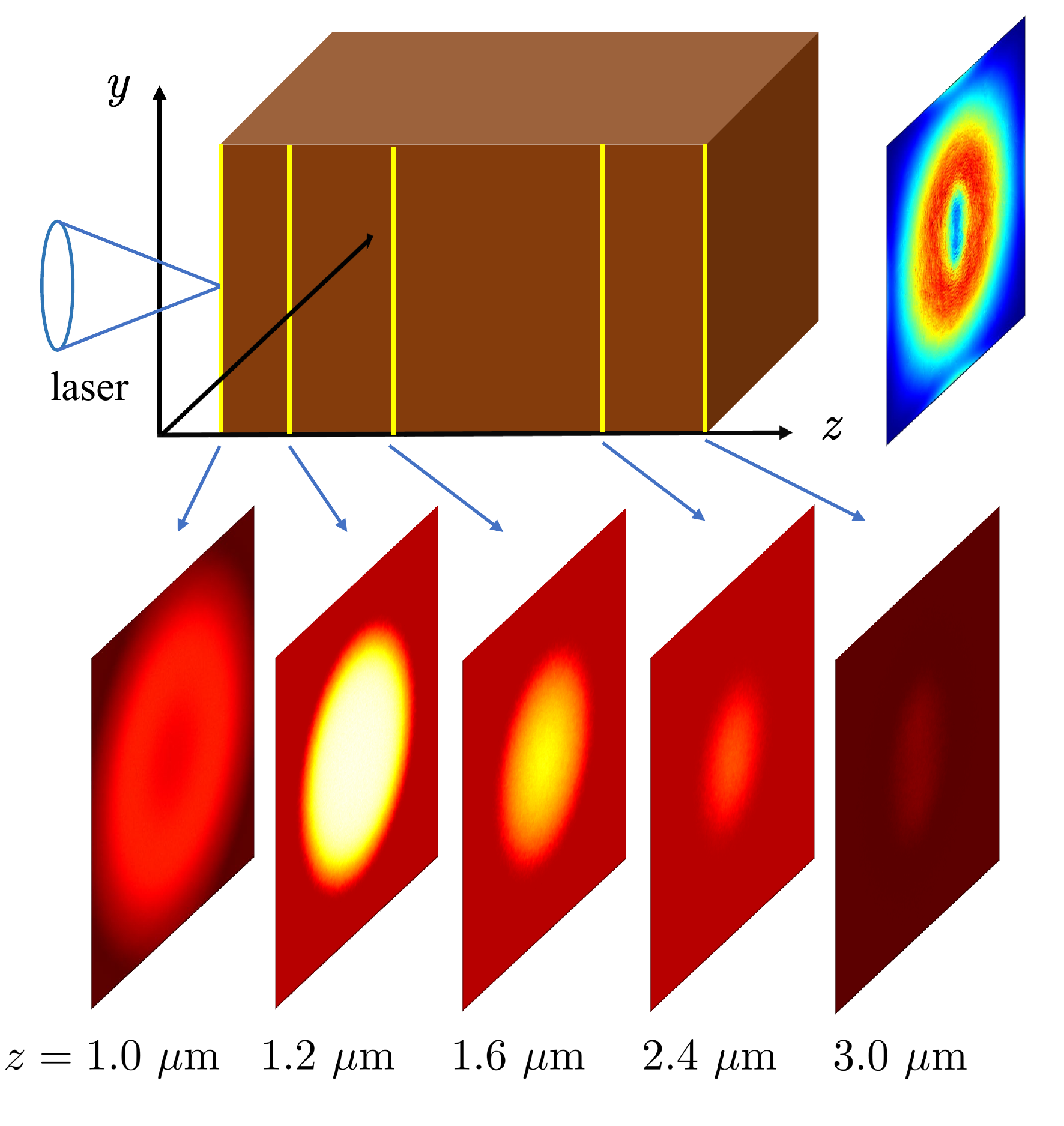} 	
	\caption{{Distribution of CTR (Upper right) and distribution of electron density in different positions of target for a realistic 3D PIC simulation containing collision and ionization effects (Bottom).} Same colorbar is used for electron density.}
	\label{fig3}
\end{figure}

\begin{figure}[!t]
	\centering
	\includegraphics[scale=0.5]{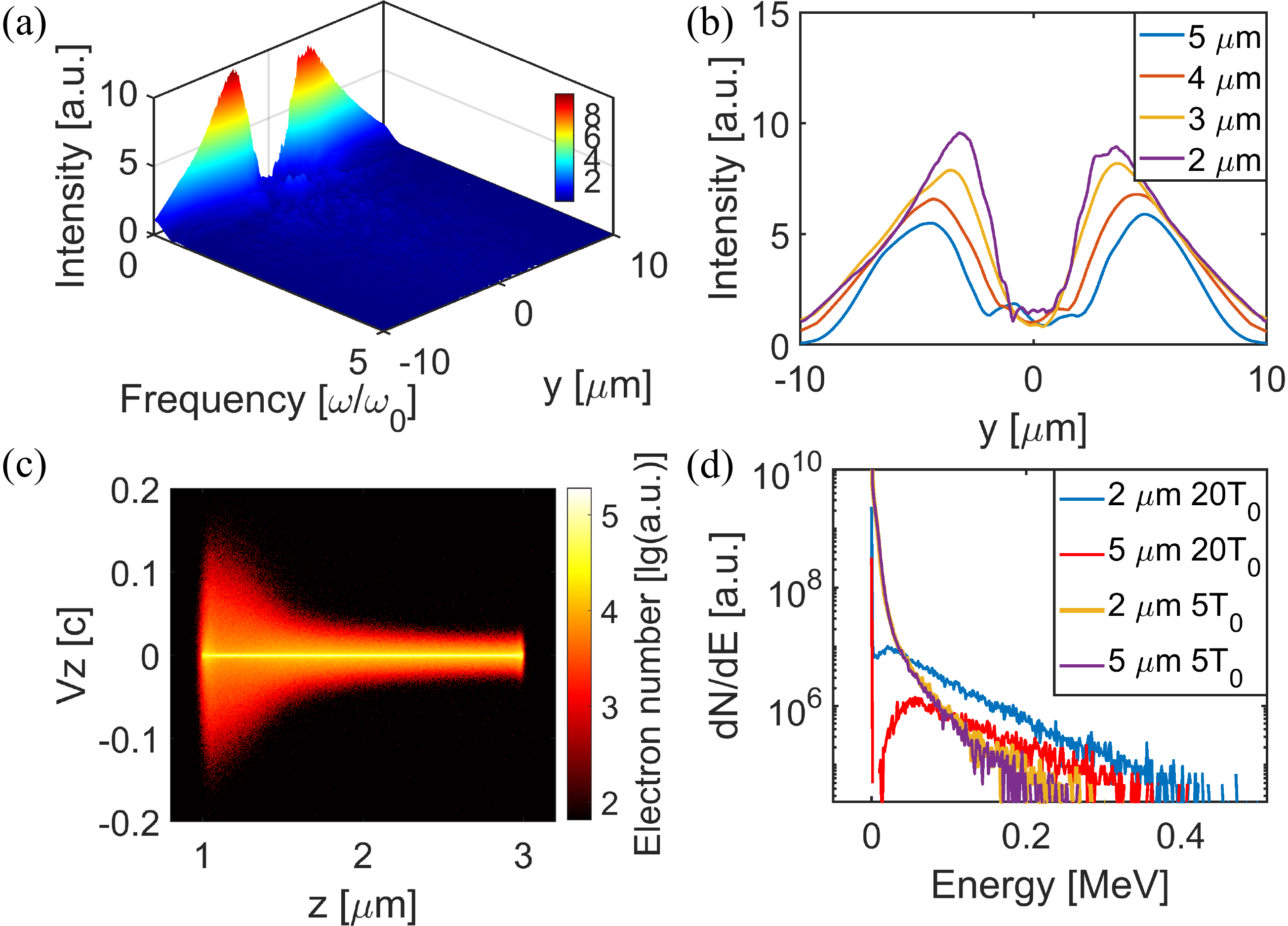}	
	\caption{{2D realistic PIC simulation results including collision and ionization effects.} (a) Spectrum of CTR for $2\ \mu \mathrm{m}$ target. (b) CTR intensity distribution, different colors represent targets with different thicknesses. (c) {Electrons phase space distribution with the thickness of target of $2\ \mu \mathrm{m}$ at $20\ T_0$, where $z=1\ \mu \mathrm{m}$ is the front surface of the target, and $z=3\ \mu \mathrm{m}$ is the back surface of the target.} (d) Energy spectra of electrons. Almost no electrons reach the rear surface of the target at $5\ T_0$.}
	\label{fig4}
\end{figure}

\begin{figure}[!t]
	\centering
	\includegraphics[scale=0.6]{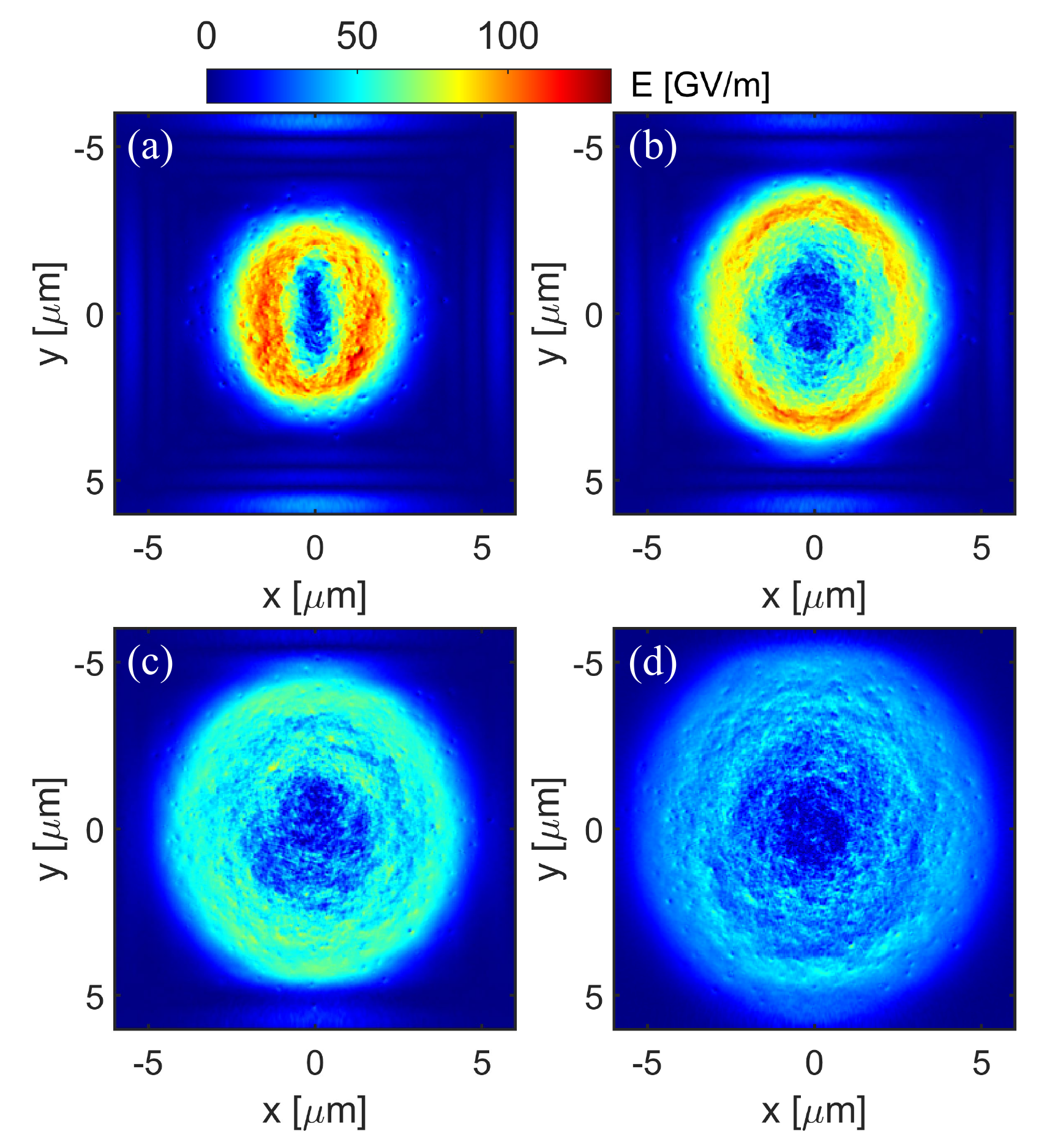} 	
	\caption{Distribution of electric field in $1 \ \mu \mathrm{m}$ behind target at (a) $10\ T_0$, (b) $11\ T_0$, (c) $12\ T_0$, (d) $13\ T_0$ for {a 3D realistic PIC simulation containing collision and ionization effects.} Same colorbar is used.}
	\label{fig5}
\end{figure}

The simulations is carried out with the PIC code LAPINS developed
by one of the authors. LAPINS is an implicit PIC code \cite{Wu2019PRE} which includes collision\cite{Wu2017PRE207}, ionization\cite{Wu2017PRE208,Wu2014POP}, radiation\cite{Wu2018hpl}, strong field QED \cite{Wu2015POP} and nuclear reaction\cite{Wu2021Aip}. Because of the amount of calculation, we decided to use a thin target for simulation and diagnose the terahertz radiation, so as to obtain the CTR distribution of 2D and 3D simulations. In our 2D simulations, the target was modeled as a uniform aluminium (Al) slab with different thicknesses. The density of the target is $2.7 \ \mathrm{g/cc}$ which is the density of solid aluminum. The initial temperature of the target is set to room temperature. The laser wavelength is $1\ \mu \mathrm{m}$ and laser radius is $3\  \mu \mathrm{m}$. The laser is vertically incident on the target and and the laser is polarized along the y direction. The laser intensity is $3.425\times10^{19}\ \mathrm{W/cm^2}$ and the duration of the laser pulse is $10\ T_0$, where $T_0=\lambda/c$ is the normalized time unit. The 2D simulations were carried out in a Z-Y Cartesian geometry which is parallel to the propagation direction of laser. The absorption field and particle boundary conditions are set in the Y and Z directions. The laser parameters of our 3D simulation are exactly the same as those of 2D simulations. We selected a target with a thickness of $2\ \mu \mathrm{m}$ for 3D simulation. {Especially, in order to be close to real situations, collision effect and ionization effect are both included in all of the simulations displayed in this subsection.} The Monte Carlo collision model in LAPINS includes binary collisions between electron-electron, electron-ion, and ion-ion\cite{Wu2017PRE207}. Besides, the ionization model in LAPINS includes the collision ionization, electron-ion recombination and ionization potential depression. This model can not only be applied for plasmas near thermal equilibrium, but also can be used in relaxation of ionization dynamics\cite{Wu2017PRE208}. After considering the collision effect and ionization effect, our PIC simulation is very close to the real situation.

Figure\ \ref{fig3}(bottom) shows the density distribution of electrons at different positions in the 3D simulation, where $z=1.0\ \mu \mathrm{m}$ and $z=3.0\ \mu \mathrm{m}$ are the front surface and the back surface of the target, respectively, and the same colorbar is used between the front surface and the back surface of the target. We can see that except on the front surface of the target, the electron distribution in other positions is not looped, and there is no ring formation in the electron beam transport process. The reason for the annular distribution of electron density on the front surface is that the target is deformed by the incident laser.  {Figure\ \ref{fig3}(upper right) shows the annular intensity distribution of CTR at the plane $0.5\ \mu \mathrm{m}$ behind the target and perpendicular to the laser propagation direction in 3D simulation. This is obtained by firstly taking Fourier transformation of the combined electric field strength in X and Y directions at the plane $0.5\ \mu \mathrm{m}$ behind target and then picking up the low frequency limit.} This plot shows that the CTR is indeed annular.

The spectrum of CTR can refer to figure \ref{fig4}(a). It is the spectrum distribution of CTR for 2D simulation at $2\ \mu \mathrm{m}$ behind target. We diagnose the electric field in the Y direction along a straight line parallel to the Y axis and do Fourier transform on it to get the spectrum. As we can see, there is a peak at low frequency and the optical transition radiation is difficult to distinguish. Therefore, its intensity distribution and spectrum characteristics is consistent with the theory. Figure\ \ref{fig4}(b) shows the intensity distribution of CTR for target with different thicknesses. As we can see, as the thickness of the target increases, the intensity of CTR decreases, and the opening angle of the cone becomes larger and larger. {This result can be well understood if the energy loss of fast electrons in the target is taken into account, resulting in different speed of the electron beam passing through the back surface of the target with different thickness. As our 2D simulation is 2D in spatial spaces however 3D in velocity spaces, and the entire collision and ionization physics are included, therefore, this result can be well extended to full 3D situations.}

To explain more accurately that the radius of the ring increases with the thickness of the target. In figure \ref{fig4}(c), we have drawn the phase space distribution of electrons. We can see that the electrons in the target have a deceleration process. Moreover, there will be a certain deceleration after passing through the sheath field generated by the front surface. In order to see the deceleration effect of the target on electrons more clearly, we show the spectra of electrons in figure\ \ref{fig4}(d). The orange and purple lines represent the kinetic energy distribution of electrons in $2 \ \mu \mathrm{m}$ target and $5 \ \mu \mathrm{m}$ target at $5\ T_0$, respectively. At this time, almost no electrons reach the rear surface of the target. We can see that the distribution of electrons produced by targets with different thicknesses is almost the same. The blue and red lines represent the kinetic energy distribution of electrons passing through the back surfaces of $2 \ \mu \mathrm{m}$ and $5 \ \mu \mathrm{m}$ targets at $20\ T_0$, respectively. From this, we can see that the electrons passing through the thick target are obviously less than those passing through the thin target, because the thicker the target, the greater the kinetic energy loss of electrons. This reason makes the thicker the target, the larger the ring of CTR diagnosed. The research of A. J. Kemp et al. and M. Sherlock et al. shows that the energy loss of electron beam in the high density target may be due to resistive heating, the collision effect, and the interaction between wave and plasma\cite{Kemp2006PRL,Sherlock2014PRL}.

In addition, because of the relatively low frequency of the CTR in our simulation, we can directly diagnose the electric field to approximately describe the behavior of the CTR. By diagnosing the distribution of electric fields in the X and Y directions at different times, we found that the circle formed by the electric field increased with time, as shown in figure\ \ref{fig5}. The (a), (b), (c) and (d) in the figure\ \ref{fig5} represents the total electric field intensity distribution in the X and Y directions at $1 \ \mu \mathrm{m}$ behind the target at the $10\ T_0$, $11\ T_0$, $12\ T_0$ and $13\ T_0$ in 3D PIC simulation, respectively. That is reasonable, because the faster electrons pass through the back surface of the target preferentially, and the cone angle of the radiation field formed by them is smaller, while the slower electrons pass through the back surface of the target latterly, and the cone angle formed by them is larger, therefore the ring of the radiation field becomes larger and larger with time. If we can measure the change of the angle of the CTR with high time resolution, we can infer the energy spectra of the electron beam near the surface of the target.

\subsection{{Artificial simulations without taking collision and ionization effects into account}}
\begin{figure}[!t]
	\centering
	\includegraphics[scale=0.6]{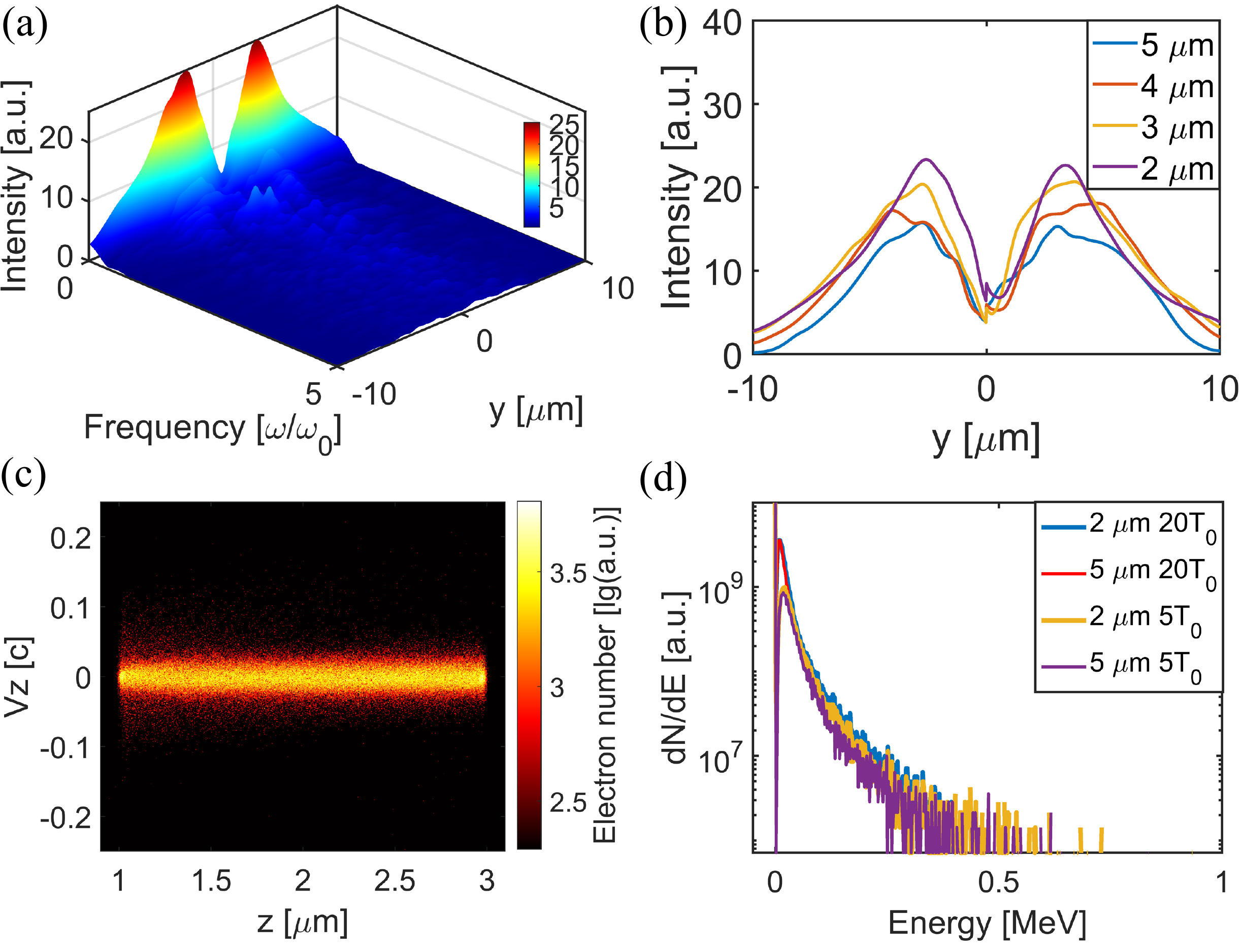} 	
	\caption{{2D artificial PIC simulation results without taking collision and ionization effects into account. (a) Spectrum of CTR for $2\ \mu \mathrm{m}$ target. (b) CTR intensity distribution, different colors represent targets with different thicknesses. (c)  Electrons phase space distribution with the thickness of target of $2\ \mu \mathrm{m}$ at $20\ T_0$, where $z=1\ \mu \mathrm{m}$ is the front surface of the target, and $z=3\ \mu \mathrm{m}$ is the back surface of the target. (d) Energy spectra of electrons.}}
	\label{fig6}
\end{figure}
\begin{figure}[!t]
	\centering
	\includegraphics[scale=0.6]{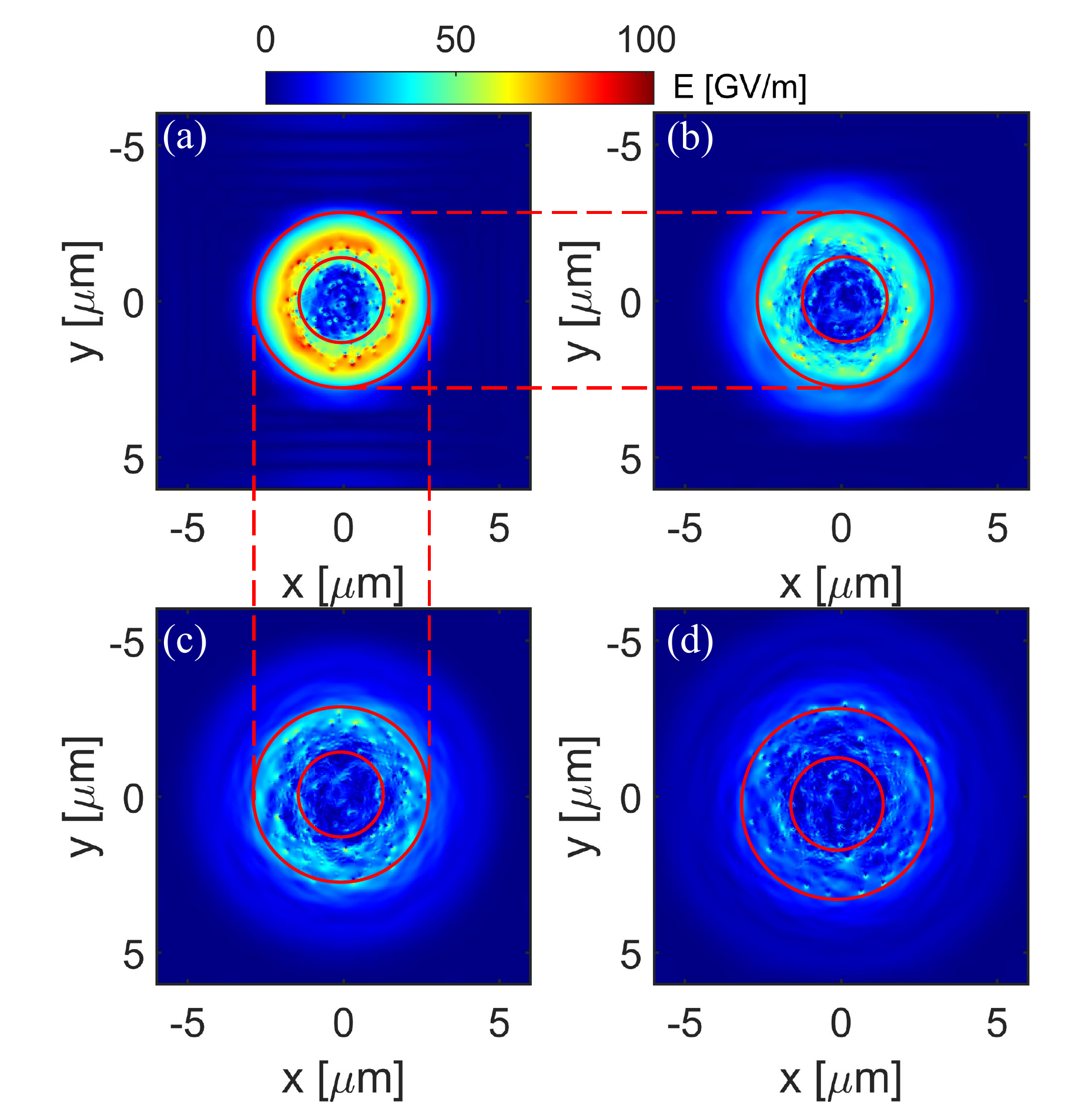} 	
	\caption{{Distribution of electric field in $1 \ \mu \mathrm{m}$ behind target at (a) $13\ T_0$, (b) $14\ T_0$, (c) $15\ T_0$, (d) $16\ T_0$ for a 3D artificial PIC simulation without taking collision and ionization into account. The area surrounded by the red ring in the figure is the area with high strength. The red dotted line is to indicate the range of the ring. Same colorbar is used.}}
	\label{fig7}
\end{figure}
{To understand the role played by collision and ionization physics, artificial simulations without these effects are performed, as with most of the previous works. The ionization state of the target in artificial simulations is set to trivalent.}
\par
{
Figure \ref{fig6} show the 2D artificial PIC simulation results without taking collision and ionization into account. Comparing figure \ref{fig4}(a) with figure \ref{fig6}(a), we can see that the spectrum shapes of CTR are the same for realistic simulations and artificial simulations. However, the intensity of the artificial simulations are higher than realistic simulations. Meanwhile, in the figure \ref{fig6}(b), the opening angles of the CTR remain almost unchanged in the artificial simulations. This is because electrons transport in the target with little energy loss when the collision effect is not taken into account. This weakens the dependence of the energy loss of the electron beam on the thickness of the target. It can also be drawn from figure \ref{fig6}(c) and (d). Besides, the intensity of CTR decreases when the thickness of the target is increased. This is because increase the thickness of the target makes the electron transport distance longer, and the number of electrons passing through the back surface of the target is smaller in the same amount of time. In the figure \ref{fig6}(c), show the electrons phase space distribution for $2\ \mu m$ target at $20\ T_0$. We can see that the velocity of electrons is evenly distributed along the radial direction in the artificial simulation, and there are also some electrons with large velocity. This also shows that the electron beam transport in the target with  little energy loss. The electrons transport distance required for thick target is longer. For thick targets, more electrons are still inside the target, resulting in the CTR intensity of the thick target in figure \ref{fig6}(b) lower. In the figure \ref{fig7}, we show the spectra of electrons. The orange and purple lines represent the kinetic energy distribution of electrons in $2 \ \mu \mathrm{m}$ target and $5 \ \mu \mathrm{m}$ target at $5\ T_0$, respectively. As we can see, for targets with different thicknesses, the energy spectra of electrons are almost identical, which reflects the electron beam transport in the
target with little energy loss. Therefore, the opening angles of the CTR in figure \ref{fig6}(b) remain unchanged.}
\par
{The 3D artificial PIC simulation results without taking collision and ionization into account are shown in figure \ref{fig7}. Figure\ \ref{fig7} show the total electric field intensity distribution in the X and Y directions at $1 \ \mu \mathrm{m}$ behind the target at the $13\ T_0$, $14\ T_0$, $15\ T_0$ and $16\ T_0$ for a 3D artificial PIC simulation. The area surrounded by the red ring in the figure is the area with high strength. The red dotted line is to indicate the range of the ring. From the results, we can also conclude that the spatial scale of CTR will increase with time. That's because the faster electrons pass through the back surface of the target preferentially. However, due to the uniform distribution of electrons in the phase space, high-energy electrons pass through the back surface of the target continuously. This makes the CTR in small opening angle region with high intensity which has been marked by a red ring in figure \ref{fig7} always exist.}
\par
{Comparing the realistic simulations results, we find that artificial simulations can also reflect the relationship between the diameter of CTR and the speed of electrons. However, ignoring the collision effect significantly underestimate the energy loss of electron beam when transporting in the target, which fail to describe the complete characteristics of CTR produced by electron beam on the back surface of the target.}
\section{Discussion and conclusion}
Most experiments failed to diagnose the annular CTR. However, in the experiment of Pakluea\cite{Pakluea2020}, they use an electron beam with an energy of MeV to irradiate the Al target, and diagnose the forward CTR. From their experimental results, we can see that when the CCD is placed on the focal plane of the focusing lens, circular CTR is diagnosed, while annular CTR is diagnosed at the position deviating from the focal plane. We can easily understand this result from the imaging theory of convex lens. {When the image is real and the CCD is placed at the image distance, the shape and size of CTR at the back surface of the target can be obtained according to the imaging law.} This result confirms CTR can be measured experimentally.
However in most experiments, CCD is placed at the focus of convex lens\cite{Storm2008RSI,Batani2006PPCF}, circular CTR is therefore diagnosed.
Full PIC simulation of solid density plasma is extremely difficult. This is because, in order to reduce numerical heating\cite{Lubos,Langdon1970JCP,Wu2019PRE,Wu2018hpl,Wang2014CPB} and suppress numerical instability, the grid size for general explicit PIC codes is of Debye length, which makes PIC simulation of sizeable solid density plasma almost impossible. In LAPINS code, a fourth-order spatial difference scheme is combined with an implicit scheme for temporal stepping in solving electromagnetic fields. This new scheme \cite{Wu2019PRE} can completely remove numerical self-heating and significantly reduce the simulation burden by using coarse simulation grids when simulating solid-density plasmas. This code enables us to calculate coupled atomic and plasma processes for intense laser–solid interaction in a more realistic way than previous codes. Within the simulations, the ionization charge state and conductivity (or resistivity) of the target can self-consistently evolve in a precise manner according to the local plasma and electromagnetic field conditions. Different types of materials can now be modeled, with account taken of their intrinsic atomic properties.

However one needs to note, although the simulation burden is significantly reduced, our 3D PIC simulation still takes more than 170,000 cpu-hours. This is an unprecedented 3D PIC simulation.
Our work figures out the relationship between CTR and electron transport properties in solid. From our simulation results, we can see that the annular distribution characteristics of CTR are closely related to the properties of electron beams. This result provides a scheme for diagnosing electron transport properties by CTR. It can be used to study the energy deposition process in inertial confinement fusion and the improvement of ion beam quality in laser driven ion acceleration.

In conclusion, {the characteristics of CTR caused by fast electrons generated by laser solid target interaction are studied through PIC simulation in two different situations. Comparing the realistic simulations, artificial simulations significantly underestimate the energy loss of electron beam when transporting in the target, which fail to describe the complete characteristics of CTR produced by electron beam on the back surface of the target. From the realistic simulation containing collision and ionization effects, it is found that the diameter of  CTR increases when the thickness of the target is increased.} This is due to the more energy loss of thick target to electron beam. In addition, we find that the radius of the ring formed by the CTR generated at fixed positions is increasing with time, which can used to infer the energy spectra of fast electrons. Our finding may be used as a new tool to diagnose the electron energy spectra near the surface of solid density plasmas.

\ack
This work was supported by the National Natural Science Foundation of China (Grant No. 12075204, No. 11875235, and No. 61627901), the Strategic Priority Research Program of Chinese Academy of Sciences (Grant No. XDA250050500) and Shanghai Municipal Science and Technology Key Project (No. 22JC1401500). Dong Wu thanks the sponsorship from Yangyang Development Fund.
\section*{reference}
\bibliographystyle{iopart-num}
\bibliography{reference}

\end{document}